\documentstyle[12pt,aasms4]{article}

\slugcomment{To appear in ApJ Supplements}

\lefthead{Garilli et al.}
\righthead{Photometric properties of clusters of galaxies}

\begin{document}

\title{PHOTOMETRIC PROPERTIES OF CLUSTERS OF GALAXIES}
\author{Bianca Garilli, Dario Bottini and Dario Maccagni}
\affil{Istituto di Fisica Cosmica del CNR, via Bassini 15, 20133 Milano, Italy}

\author{Luis Carrasco\altaffilmark{1} and Elsa Recillas\altaffilmark{1}}
\affil{Instituto Nacional de Astrofisica \'{O}ptica y Electr\'{o}nica, Apartado Postal 51 y 216, 72000 Puebla, Pue., Mexico}

\altaffiltext{1}{Also Instituto de Astronomia, UNAM, Mexico D.F., Mexico}

\begin{abstract}

We report the first results of a
long term program aimed at investigating the photometric
properties of the cores of Abell and X-ray selected (EMSS,
\cite{emss}) clusters of galaxies. 
We observed 77 clusters of galaxies in the redshift range $0.05\leq z \leq 0.25$ 
in the Gunn $g$, $r$ and $i$
filters: in this paper we present the data on 59 clusters with
good absolute photometry and on another group of 8 clusters  with
acceptable relative photometry. For all these clusters we
show color--magnitude diagrams in the two colors: when, as 
in most cases, the early type galaxy sequence is identifiable, we
compare it with the expectation from the Virgo c--m relation (\cite{vs}) 
and find that the Virgo relation holds to $z\sim0.2$.
We do not find any sign of active evolution in cluster galaxies
since that epoch, nor in the
percentage of blue galaxies or in the early type galaxy colors,
if we accept that the scatter is of the order of 0.2--0.3 mag
with respect to the expectations on the basis of standard k--corrections.
We point out the presence of a certain number of anomalously red
galaxies in the $r-i$ color, which are too red with respect to
their $g-r$ color to be normal field galaxies at a redshift higher
than the cluster one. Finally we briefly compare a few properties
of the two subsamples of optically selected
and X--ray selected clusters.

\end{abstract}


\section {Introduction}

The fact that large numbers of coeval galaxies can be found only in clusters
enhances the importance of these objects as tools in cosmological studies,
at the same time providing the substantial database needed to investigate
galaxy evolution in a controlled environment. Galaxy clusters can be
catalogued and observed up to $z\sim1$, thus providing a reasonable time span 
over which the main observable galaxy properties can be compared. 
\par
Up to the end of the '70s,
clusters of galaxies were optically selected via the
galaxy density contrast over the field (\cite{a1,zw,aco}, 
ACO). This favors the selection
of rich compact clusters, both at low and high redshifts.
The Extended Medium Sensitivity
Survey (EMSS, \cite{emss}), based on the {\it Einstein Observatory}
data, has proved that complete catalogues of
clusters of galaxies can be obtained by detecting
the X-ray emission of the hot intergalactic medium.
In the X-rays, cluster visibility 
is  based on the effective 
presence of a potential well in which the hot gas is confined {\it and} 
on the physical properties of this gas (temperature and density).
The X-ray properties of
clusters have been widely studied (e.g., \cite{b81,es}), and their X-ray properties correlated with
optical properties, to seek a correspondence between
properties of the gas (e.g. temperature, luminosity) and
properties of the luminous matter (galaxy density, distribution, etc.).
In principle the two selection methods are capable of detecting
the same class of objects, as shown by a number of Abell clusters 
found in the EMSS,
in practice, however, several biases are introduced, depending on 
morphology, richness, and the scatter in the relations between 
optical and X-ray properties.
It is well known
(\cite{sca}) that even the ACO catalog
is incomplete for $z$ approaching 0.2 and it is biased toward rich objects.
Samples drawn from both optically selected {\it and} X-ray selected catalogs
could minimize the overall biases and contain clusters of galaxies spanning the whole range of
cluster masses and morphologies, from which to derive the
properties of the luminous matter and the forms of its aggregation.
At present, mixed samples with reasonable numbers of clusters
can be constructed only up to $z<0.2-0.3$, which corresponds approximately to the
look back time when evolutionary phenomena, like the Butcher--Oemler effect
(\cite{bo78,bo84}) become detectable.
\par
To fully
understand evolutionary phenomena, it
is essential
to have a solid and statistically significant knowledge of the properties
of clusters and of their galaxies for the local universe, where by local
we mean a volume where we can reasonably think that galaxies have
undergone only passive evolution, i.e. where
the Butcher--Oemler effect is negligible. 
Spectroscopic observations are needed to study the cluster dynamics 
and to have a better membership probability, but to reach fainter magnitudes, even using a multislit
spectrograph, a large amount of time with medium size
(2-4 m) telescopes is needed.
On the other hand, by limiting the photometry to the cluster core
for reasonably nearby clusters (in a redshift range from $z\sim0.05$ to 
$z\sim0.2$) the contamination by non cluster galaxies is expected to be
less than 10\% and can be accounted for statistically.
\par
In this paper we present the multicolor photometric data
of a sample of clusters which follows the rationale outlined above.
We first illustrate the sample composition, observations and
data reduction procedures and establish the photometric accuracy. 
Then we show the color--magnitude
diagrams which are then used to isolate the blue and red galaxies in the cluster cores.
\par
Throughout this paper we adopt $H_0 = 50$ km~s$^{-1}$~Mpc$^{-1}$ and
$q_0=0.5$.

\section{The Data}
\subsection{The Sample}

The sample we observed comprises: (a) 39 Abell clusters; (b) 21 EMSS clusters; 
(c) 7 EMSS clusters which are also included in the ACO catalog.
ACO and EMSS clusters have been selected in such a way to mantain a similar redshift distribution.
Furthermore, 
Abell clusters cover all morphological types (according to the
Bautz--Morgan classification as reported in the ACO catalog) and richness
classes. We secured the redshifts for two clusters previously at unknown
distance 
(spectroscopy of galaxies in the cluster sample will be presented in
another paper). We are thus able to present data on 67 clusters with
measured redshift.
\par
Figure 1 a to d gives a pictorial representation of the known properties of our 
cluster sample: the redshift distributions of the 39 Abell clusters and of
the 21+7 EMSS clusters; the X--ray luminosity distributions of the
17 Abell clusters observed by the {\it Einstein Observatory}
and of the 21+7 EMSS clusters; the Bautz--Morgan type and the richness (as
defined by the C parameter in ACO)
distributions of the Abell clusters. The X--ray luminosity of the Abell clusters
has been 
computed from the
IPC count rate, as obtained from the {\it Einstein Observatory Catalog
of IPC X--Ray Sources}, assuming a 
Raymond--Smith spectrum with a temperature of 6 keV, as it was 
done in the EMSS.
\par 
Table 1 reports this same information in unbinned form for all the
67 clusters in the sample. Column 1
gives the cluster name (either the Abell number or the EMSS truncated
source coordinates or both), columns 2 and 3 give the catalog cluster
center coordinates for the epoch 1950.0, column 4 gives the redshift:
the two new redshifts so far unpublished are marked with an asterisk,
while redshifts for the other Abell and EMSS clusters are taken from \cite{aco} and
from \cite{stocke} respectively; columns 5 and 6  the 
Bautz--Morgan type and the number of galaxies between $m_3$ and 
$m_3+2$ as reported 
in ACO (the C parameter) respectively,
and finally in column 7 we give the X--ray luminosity, in units of 
10$^{+43}$ ergs s$^{-1}$.

\subsection{The Observations}

The cluster multicolor photometry was carried out in the
years 1990--1994 at the 2.1m telescope of the
Mexican National Astronomical Observatory at San Pedro Martir, 
equipped with a CCD of 384x576 23$\mu$ pixels,
resulting in a field of view of $\sim$ 2x3 arcmin.
Since December 1993, a new 1024x1024 22$\mu$ pixels CCD 
has been available,
thus 12 clusters have been observed with a larger field 
of view ($\sim$ 4x4 arcmin). For 10 clusters we made two 
or more pointings, to
explore a larger area. The total number of nights allocated to
this program has been 34, 18 of which proved to be photometric. 
All clusters have been observed through the three {\it g, r} 
and {\it i} Gunn filters with typical
exposure times between 20 minutes and 1 hour per filter. When
the total exposure time was greater than 20 minutes per filter,
the observation was splitted into 2 or 3 exposures,
to reduce cosmic ray contamination. 
The average seeing was $\sim$ 2 arcsec.
\par
Standard
stars in the Gunn filters (taken from the lists of \cite{tg} and \cite{wade}) 
were observed all through the nights between
cluster exposures
to derive flux calibrations. During photometric nights,
the accuracy in the zero point is of the order of 0.05 mag.
Some clusters were observed 
in conditions which turned out to be non-photometric. In these cases shorter exposures
on the same target were successively made in good weather conditions, 
and images have been calibrated by matching the magnitudes of
stars or compact galaxies in the field.  
\par
The journal of the observations is given in Table 2, for the
67 clusters in the sample.
In column 1, the cluster name is given (an asterisk marks clusters
observed in non-photometric conditions), 
in column 2 the total area
explored in Mpc$^2$ at the cluster redshift, in column 3 the number of galaxies
detected in all three filters. Columns 4 to 15 contain the information pertaining to
each filter separately: filter id, the exposure time in seconds,
seeing in arcsec, and the limiting magnitude.
Such limiting magnitude has been computed for each image,
and represents the fainter apparent magnitude a pointlike
source can have to be detected with a signal to noise ratio
of three. The
images generally allow to detect objects as faint as 22.5 {\it r}
mag in all the three filters. 

\subsection{Data Reduction}

Data reduction was performed in a standard way: 
after the usual bias subtraction, flat fielding and cosmic ray removal, 
images in the same filter have been registered and summed.
\par
By careful inspection of the images, we produced lists of objects for each
filter. These lists were complemented with the ones produced by the
MIDAS/INVENTORY search algorithm run with a low detection threshold and
crosschecked among the three independent images. Only objects detected
in all the three filters were retained in each cluster catalog. As
a second step we rejected all objects which were below any of the
limiting magnitudes reported in Table 2. This procedure does not produce
clear--cut magnitude limited catalogs, because the effective limits
are dependent on the colors of the objects, but exposure times were such
that normal ellipticals of $M_r=-18$ are generally included in the
catalogs of galaxy clusters at $z=0.2$. 
\par
Finally, bright stars have been removed from
the catalogs. Convolution
of a Point Spread
Function, representative of our images, with exponential and de Vaucouleurs'
galaxy profiles have shown that, beyond the complexity of the two parameter
dependence, galaxies could be separated from stars if the objects
were brighter than $m_r\sim21$ (\cite{pocar}). Thus bright stars
were removed from the catalogs on the basis of their FWHM. 
We recall that in the field,
stars are a factor 2.5
less numerous than galaxies at $m_r=22$
(see e.g. \cite{tyson}). 
\par
Two kinds of magnitudes have been computed for each galaxy:
an apparent aperture magnitude, and a metric magnitude. 
Metric apertures are computed within 10 kpc radius.
Magnitudes, and errors, have
been computed using the IRAF package {\it apphot}. 
In deriving galaxy magnitudes and colors, we have considered
both seeing effects and the presence of nearby objects. 
To include most of the signal in the aperture, angular
diameters have been chosen to be twice as large as the largest seeing 
FWHM in the three filters.
In the cases when
the aperture magnitude is strongly contaminated
by nearby objects, the 10 kpc magnitude has been used. When
both magnitudes were affected by nearby objects, 
the smallest aperture has been used, but a minor weight was
given to the point.
Magnitudes have been corrected for atmospheric extinction
and galactic absorption on the basis of the
measured hydrogen column density along the
line of sight (\cite{stark}), converted to A$_g$,
A$_r$ and A$_i$ following \cite{sgh}.
Statistical errors on magnitudes (and colors) depend on the
signal to noise ratio,and therefore also on apparent magnitude.
In Table 3, we illustrate the statistical accuracy
of the color determination as a function of magnitude for the 
whole sample:
90\% of the galaxies brighter than  m$_r$=21
have errors on $g-r$ smaller than 0.068 mag and on 
$r-i$ smaller than 0.058 mag; 80\% of the galaxies brighter than  
m$_r$=22 have errors on $g-r$ smaller than 0.09 mag
and on 
$r-i$ smaller than 0.076 mag.
\par
The number of galaxies in each cluster field for which we have 
multicolor information ranges from 13 to 175. Colors are 
available for a total of 3843 galaxies.  

\subsection{Photometric Accuracy}

To check our photometric accuracy, we have compared our magnitudes 
with those obtained
by \cite{hs} (HS), who observed 185 Brightest 
Cluster Galaxies in
the {\it g} and {\it r} Gunn filters. For this comparison, we
use $H_0 = 60~km~s^{-1}~Mpc^{-1}$ and magnitudes within 
16 kpc radius apertures, as they did. No galactic extinction was applied,
as done by  HS.
For the 12 galaxies in common with HS,
in Figure 2 a and b we plot our magnitudes vs. theirs. 
The dotted line represents the
$y=x$ relation. A linear regression performed on the data of 
Figures 2 a and b
give results that are compatible with the $y=x$ relation, although 
a slight shift
is present in the {\it g} magnitudes, in the sense that 
$<g_{this work}> = <g_{HS}> -0.20
\pm 0.14$ (the same relation for the {\it r} filter is
$<r_{this work}> = <r_{HS}> -0.004
\pm 0.13$). Such discrepancy could be due to the slightly different filter+detector
response we have in the {\it g} band. We note however that the offset is 
comparable to the statistical and systematic errors one incurs in 
when computing 
magnitudes of large extended objects like the BCGs are.
No comparison can be made for the {\it i} filter, for lack of other observations. 

\section{Results}

\subsection{Color--magnitude Diagrams}

In Figure 3a we show the color magnitude diagrams for all the 59
clusters within our sample observed in optimal photometric conditions, 
sorted in increasing redshift:{\it g-r} and {\it r-i}
vs. the metric aperture in the {\it r} filter. Figure 3b shows the same
diagrams for the 8 clusters which have been recalibrated.
Error bars on magnitude and colors are indicated: they are smaller
than the point size for all but the faintest objects.
The superimposed dashed line represents the expected color-magnitude relation as
derived from \cite{vs}. The slope of the c--m
relation has been computed for our two colors by directly reading from their Figure 2 the values corresponding
to the effective wavelengths of our filters + CCD system. We then assumed
rest frame colors of $g-r=0.37$ and $r-i=0.31$ for a galaxy of absolute
magnitude $M_r=-23$. These rest frame colors are the ones obtained by
\cite{sgh} in their study of cluster BCMs, once we allow for a 0.1 shift in
$g-r$, in order to align our $g$ magnitudes with those of HS. The k--corrections
of SGH were used to convert from rest frame colors to the
colors expected at the cluster redshifts.
\par
Most clusters show a neat sequence of early type galaxies in both colors. 
Exceptions are A468 and MS1125+4324 and perhaps MS0013+1558, 
MS1154+4255
MS1401+0437 and MS1308+3244, 
to which we must add A439. Lack of an identifiable sequence
can occur when the morphological composition is strongly biased toward late types
or if the observed region is not the cluster center. 
This second occurrence
is to be excluded in the case of the X--ray selected clusters, since the
available coordinates are precise enough to identify the cluster center.
\par
In the following we will consider only the 59 clusters with top quality photometry 
(Fig. 3a). From the plots, it is clear that: a)
the slope of the c--m relation describes
quite well the data for all clusters with an identifiable sequence. That is to say, 
the c--m effect does not show any evolution
since $z\sim0.2$. b) The adopted colors and absolute magnitude normalization is
quite satisfactory for most clusters, but there are cases when 
one or both colors appear either too blue or too red, beyond any
(systematic) photometric error. In Table 4 we list the clusters
for which the difference between the "expected" c-m normalization
and the observed one differs by more than 0.1 mag in one of the two colors.
As shown in Table 3, a shift of 0.1 in color is well out of the statistical errors, 
much more so if we take into account that galaxies 
determining the early type sequence are mostly brighter than
m$_r\sim$20.5.
\par
It is worth noting that a difference in color of 0.1 cannot be ascribed 
to errors in the determination of the galactic absorption.
For instance, as far as $g-r$ is concerned (this color 
is much more sensitive
than $r-i$ to the interstellar reddening), a 0.1 shift 
would imply an offset of the same order in E(B-V), which translates into
$10^{21}$ atoms cm$^{-2}$ in terms of $N_H$. Such values are 
well above
the uncertainties quoted by \cite{bh} and/or \cite{stark}. 
We have also checked whether discrepancies of this order
exist between the
measures of \cite{stark} and \cite{bh}. Only in the case of 
A478 the two measures showed such a large difference, as it is discussed
below.
\par
We will now discuss the few cases worth particular notes.

$A272$. This is a mosaic of two fields, one of which had to be
recalibrated. The high dispersion in the sequence is probably due
to the lower photometric accuracy.

$A478$. The early type galaxies in this cluster are much redder than
expected, both in $g-r$ and in $r-i$. As noted by \cite{bs77},
this cluster is heavily reddened (they estimate a $E(B-V)=0.7\pm0.2$).
The colors of the early type galaxies in this cluster would become
as expected from the Virgo c--m relation if the hydrogen column density
is more than a factor of 3 higher than the value interpolated from
the Stark et al. catalog, i.e. if $E(B-V)\sim0.4$. This is the only case 
where the value for E(B-V) found in \cite{bh}
survey is significantly different from the \cite{stark} one:
0.24 against 0.1. Even the highest value falls a factor
of 2 short of what would be needed to bring
A478 colors to "normal". However, the different values in the two
catalogs could indicate
 that the absorption in the direction of A478 is extremely patchy 
(the closest Stark et al.'s measurement
is 0.44 degrees away from the position of the brightest galaxy
in A478).

$A175$. Galaxies in the diagram come from 4 adjacent fields. The
brightest galaxy is a foreground spiral. The scatter in the sequence is
rather large.

$A180$. The sequence is not so well defined as in other clusters and it
might be slightly bluer in $g-r$ than expected.

$A612$. The redshift of this cluster is based on the spectrum of
one galaxy we obtained in November 1994 and should be considered
provisional.

$A1081$. This cluster seems bluer than expected in $g-r$. We would take
this fact with caution because the $r$ image has been recalibrated.

$A115$. The brightest galaxy in the diagram is a foreground elliptical. 

On the whole, $\sim30\%$ of the clusters have early type sequences which
are either bluer or redder than expected on the basis of our normalization
of the c--m relation. 
This result could be due to
an intrinsic scatter in the colors of early type galaxies belonging to
different clusters, similarly to what found by HS for BCMs. 
On the other hand, within the same cluster such scatter is 
considerably reduced, as shown by the color-magnitude diagrams. 
In other words,
different clusters evolve in a different way, and this is 
reflected by the different normalization of the color-magnitude relation.
If true, some other difference related to cluster evolution
should be present. We have checked if the ``bluishness'' or ``redness''
of the c-m relation is somehow tied with other cluster properties,
like Bautz-Morgan type, richness, X-ray luminosity or central galaxy 
density. The low number of objects for which this information is available
(16) does not
allow us to see any statistically significant correlation. 
The only trend which can be inferred is with redshift,
in the sense that clusters showing a ``bluer'' c-m relation are at 
redshift higher than 0.13. An exception to this rule is MS0904+1651
(A744), which is the only nearby cluster having a bluer c-m relation.
However, this trend, if real, does not hold in the opposite way,
i.e. higher redshift clusters are not always bluer than expected.

\subsection{Blue and Red Galaxies}

We defined as blue galaxies the ones with a $g-r$
color bluer by 0.3 mag than the early type sequence of the cluster.
Clusters have been subdivided into
redshift bins of 0.05, and 
the percentage of blue galaxies in each bin has been computed
over the number of galaxies brighter than $M_r=$-18.
The percentage of blue galaxies 
ranges from $\sim8$\% to $\sim13$\%, showing an increase with
redshift, although with large error bars.

An interesting feature of our color--magnitude diagrams is the
presence of a rather large number of red galaxies, defined as
those galaxies having both $g-r$ and $r-i$ colors redder than the
early-type sequence in each cluster by 0.3 mag. There are 149
such galaxies in our fields, and their color--magnitude diagrams
and color--color diagram are shown in Figure 4. The $g-r$
vs. $m_r$ diagram is quite compatible with what expected from
field galaxies in the cluster backgrounds (see, e.g., \cite{nw}): 
such a diagram would indicate a population of 
field galaxies which should be at a redshift of about 0.3-0.5.
On the other hand, there are more than 50 galaxies occupying the 
upper part of the color-color diagram. These galaxies
have magnitudes $19.5\leq m_r \leq 22$ and have $r-i>1.0$
and $0.8\leq g-r \leq 2.0$. Normal, high redshift field galaxies are not
expected to be found in that part of the diagram.
The $r-i$ color would yield a redshift $z>0.7$, but their $g-r$ color 
is not red enough if these galaxies are far away. The anomaly
of the spectral energy distribution of these objects results from their
emission in the redshifted $i$-band, i.e. above 0.7-0.8 $\mu$.
Other authors 
observed 
galaxies with similar characteristics in other cluster fields: 
in a small sample of IR selected clusters, \cite{as}  find galaxies 
with extreme infrared colors, 
unmatched by the observations of field galaxies. In less recent years, 
\cite{blw,couch,ellis} mentioned the presence of anomalously red galaxies 
in the clusters they observed. Our observations suggest that the phenomenon
is more common and requires redshift determinations 
to be
properly assessed.

\subsection{Optical vs. X-ray Selected Clusters}

Our mixed sample contains objects found over the whole sky
(the Abell clusters) and objects serendipitously detected over a
much smaller and unconnected area (the EMSS clusters). However, the photometric
properties of the cores of both classes of objects, from the slope of 
the color-magnitude diagrams, to the presence of an early-type sequence,
to the galaxy number-density gradients, do not allow one to distinguish
the origin of the cluster studied. The reasons why the EMSS
clusters are not included in the Abell catalog do not
depend on the properties,
numbers or colors of the galaxies within a few hundred kpc radius of
the centre. The cores of clusters are alike, irrespective of the selection method.

\section {Conclusions}
\par
Our two--color photometric survey of a sample of clusters of
galaxies ranging in redshift from $z\sim0.05$ to $z\sim0.23$ shows
that:
\par
(a) the photometric properties of the galaxies in the cores of X--ray selected and
optically selected clusters are substantially the same;
\par
(b) the color--magnitude effect found in Virgo generally fits well
the data for clusters in the whole redshift range we surveyed, once
appropriate k--corrections are applied. There are clusters which
deviate by 0.2--0.3 mag from the expectation, either toward the red
or the blue. This is of the order of the scatter present in the
colors of bright cluster ellipticals. The fainter early type galaxies
share the same colors, and follow the Virgo c--m effect with a scatter
which can be evaluated to be around 0.1 mag in our colors;
\par
(c) on average, there is no detectable  Butcher--Oemler type evolution 
of galaxies since $z\sim0.2$. However, the cluster to cluster
scatter of the percentage of blue galaxies in all redshift bins (even
normalized to the surveyed area of each cluster) is very
large.
\par
(d) in our fields, we found a sizable number of galaxies redder in both
colors than the early type sequence. About a third of them have anomalous
colors, in the sense that their $r-i$ color is too red to be compatible
with their $g-r$ color if they are background field galaxies.

\acknowledgements

We thank S. Pocar for help in
the reduction of some of the data. 
Discussions with G. Chincarini and his suggestions have been
invaluable in assessing several of the points we addressed in
this paper. 
We are grateful to the UNAM--OAN Time
Allocation Committee for the generous support given to this program throughout
the years. Finally, we warmly acknowledge the assistance received by all the
OAN staff both in San Pedro and in Ensenada.

\newpage

\newpage

\newpage

\begin{deluxetable}{lccccccc}
\tablecolumns{8}
\tablewidth{0pc}
\tablenum{3}
\tablecaption{Accuracy of galaxy colors}
\tablehead{
\colhead{} & \multicolumn{3}{c}{$\sigma_{g-r}$} & \colhead {} & 
\multicolumn{3}{c}{$\sigma_{r-i}$} \\
\cline{2-4} \cline{6-8} \\
\colhead{m$_r$} & \colhead{$\leq$0.05} & \colhead {$\leq$0.07} & \colhead{$\leq$0.09} & \colhead{} & \colhead{$\leq$0.05} & \colhead {$\leq$0.07} & \colhead{$\leq$0.09}} 
 \startdata
$\leq$21	& 81\%	& 91\%	& 95\%	&& 87\%	& 94\%	& 97\%	\nl
$\leq$22	& 59\%	& 71\%	& 80\%	&& 66\%	& 77\%	& 86\%	\nl

\enddata
\end{deluxetable}

\newpage

\begin{deluxetable}{lcccc}
\tablenum{4}
\tablewidth{0pc}
\tablecaption{"blue" and "red" clusters}
\tablehead{
\colhead{Name} & \colhead{$g-r$} & \colhead {$\Delta m$} &
\colhead{$r-i$} & \colhead {$\Delta m$}} 
 \startdata

A2092		& \nodata 	&	& red 	& 0.1		\nl
A744		& blue    	& 0.1	& \nodata	& \nodata	\nl
MS0301+1516	& red		& 0.2	& \nodata	& \nodata	\nl
A410		& red		& 0.2	& \nodata	& \nodata	\nl
A478		& red		& 0.2	& red		& 0.1 	\nl
A403		& red		& 0.1	& \nodata	& \nodata	\nl
A180		& blue	& 0.1	& \nodata	& \nodata	\nl
A612		& blue	& 0.1	& \nodata	& \nodata	\nl
A1081		& blue	& 0.2	& \nodata	& \nodata	\nl
MS0433+0957	& \nodata	&	& red		& 0.1		\nl
A588		& blue	& 0.1	& \nodata	& \nodata	\nl
A750		& red		& 0.1	& \nodata	& \nodata	\nl
MS0026+0725	& blue	& 0.2	& \nodata	& \nodata	\nl
A115		& blue	& 0.1	& \nodata	& \nodata	\nl
A520		& blue	& 0.2	& blue	& 0.1		\nl
MS0109+3910 & blue	& 0.2	& blue	& 0.1		\nl

\enddata
\end{deluxetable}

\newpage


\figcaption[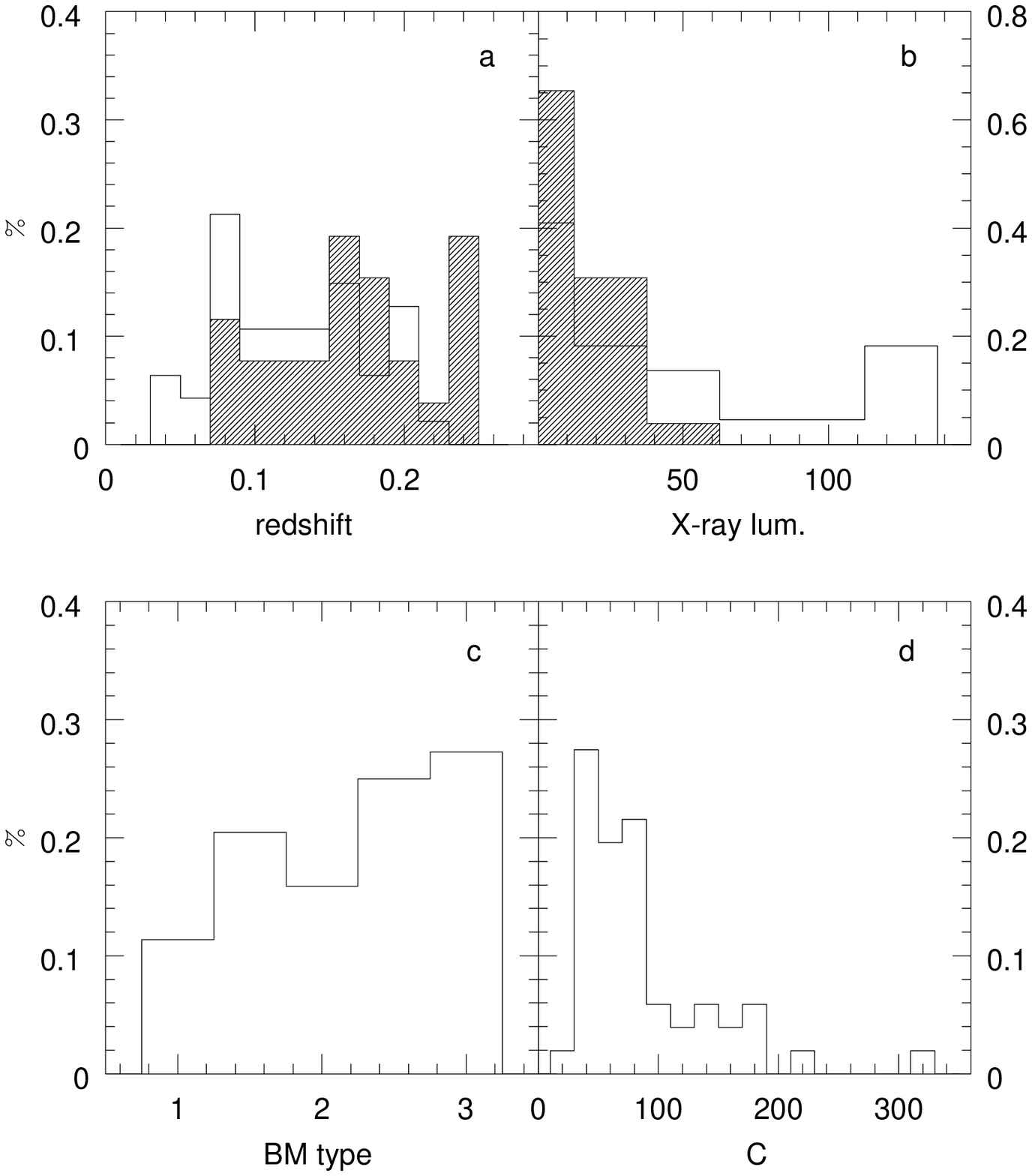]{a) redshift distribution of Abell (white) and 
EMSS (grey) clusters in our
sample; b) X-ray Luminosity distribution, symbols as in {\it a}; 
c) Morphology
distribution of Abell clusters in our sample; 
d) Richness distribution of Abell clusters in our sample}

\figcaption[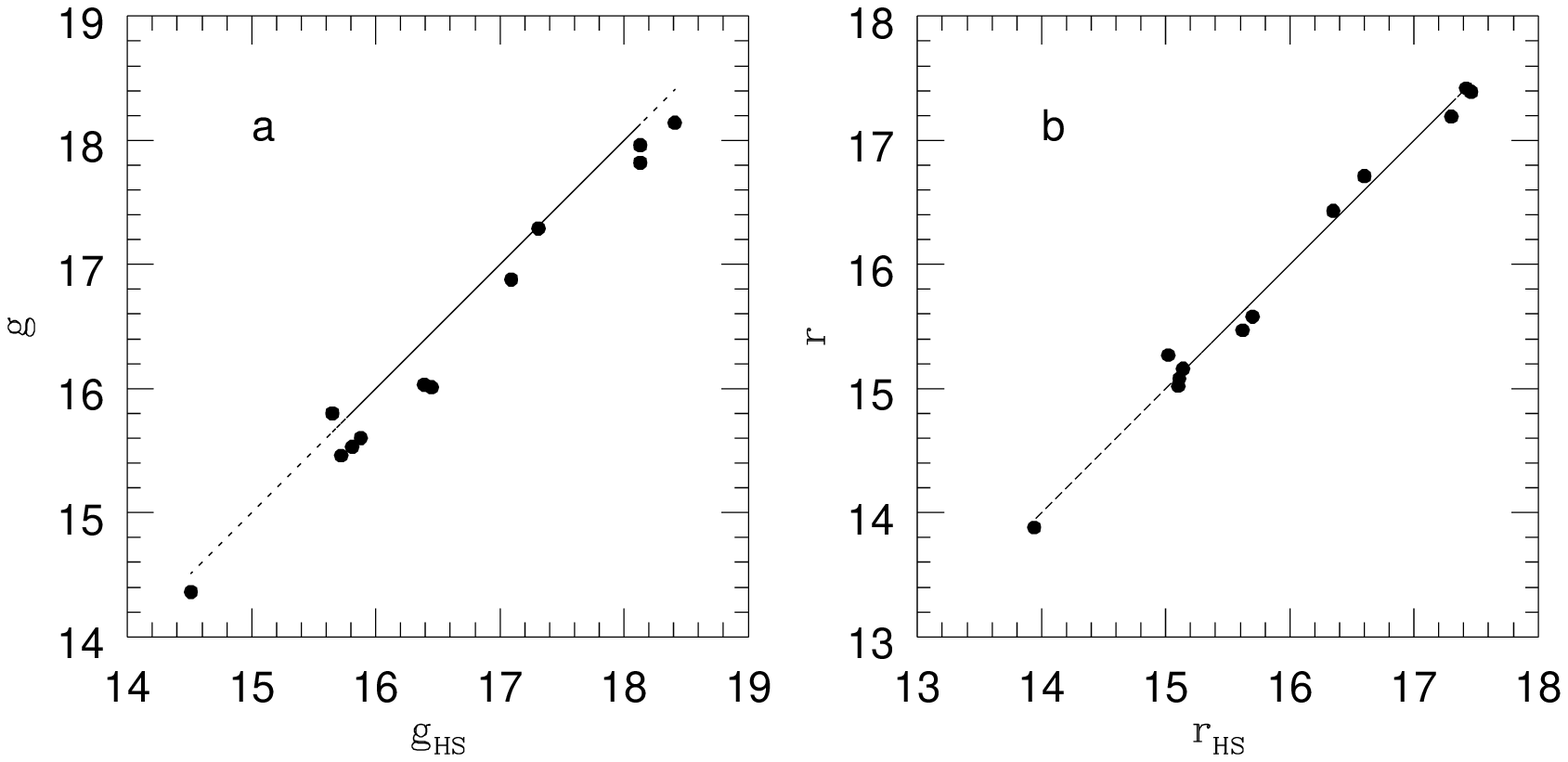]{a) our {\it g} central magnitudes vs. 
Hoessel \& Schneider (1985) magnitudes; 
the dotted line represents the $y=x$ relation; 
b) as {\it a} for the {\it r}
filter.}

\figcaption[fig3.ps]{Color-magnitude diagrams of the clusters in the sample. The
dotted line represents the virgo c-m relation (see text). a) the 59
clusters with top quality photometry; b) the 8 clusters
which have been recalibrated}

\figcaption[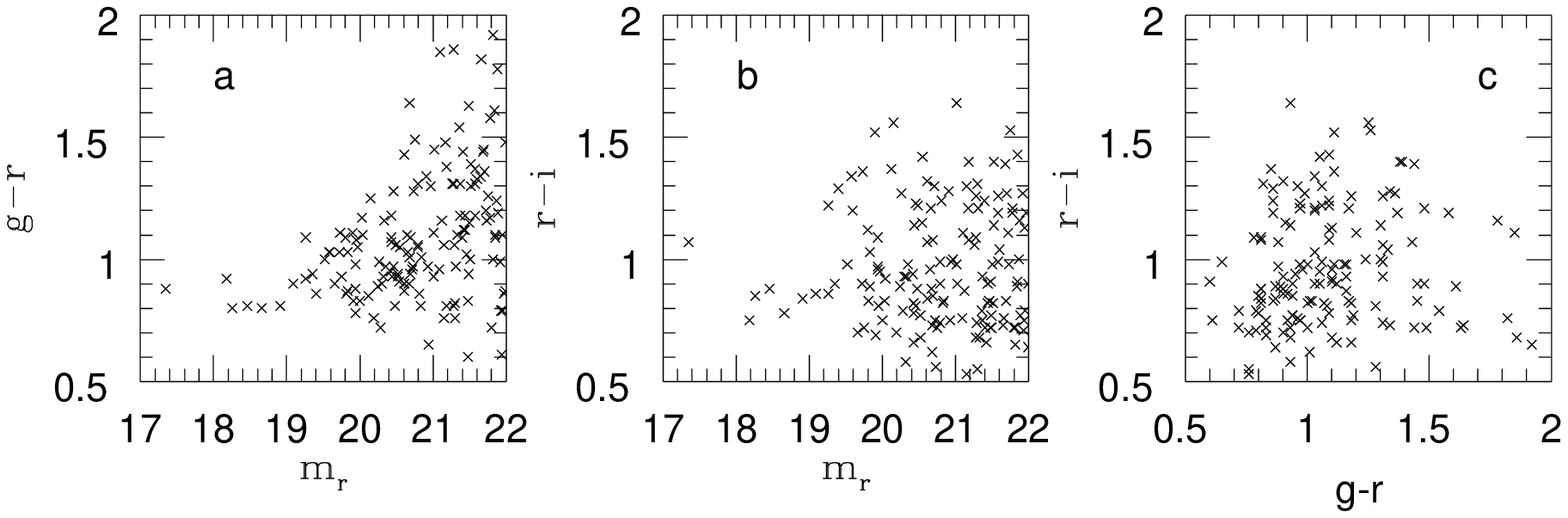]{color magnitude ($g-r$ (a) and $r-i$ (b)) and color color
diagrams (c)
of the 149 galaxies defined as "red"}

\end{document}